\documentclass[a4paper]{article}
\usepackage{ISCSLP2024}
\usepackage{ifthen}
\newboolean{blind}
\setboolean{blind}{false} 
\title{StreamAAD: Decoding Spatial Auditory Attention with a Streaming Architecture}
\name{
	\ifthenelse{\boolean{blind}}{Anonymous to ISCSLP}
	{Zelin Qiu$^{1,2}$, Dingding Yao$^1$, Junfeng Li$^{1,*}$\thanks{*Corresponding author}}
}
\address{
  \ifthenelse{\boolean{blind}}{Anonymous to ISCSLP}
  {
  	$^1$Institute of Acoustics, Chinese Academy of Sciences\\
  $^2$University of Chinese Academy of Sciences
  }
}

\email{
	\ifthenelse{\boolean{blind}}{Anonymous to ISCSLP}
	{\{qiuzelin, yaodingding\}@hccl.ioa.ac.cn, junfeng.li.1979@gmail.com}
}

\begin{document}

\maketitle
\begin{abstract}
In this paper, we present our approach for the Track 1 of the Chinese Auditory Attention Decoding (Chinese AAD) Challenge at ISCSLP 2024. Most existing spatial auditory attention decoding (Sp-AAD) methods employ an isolated window architecture, focusing solely on global invariant features without considering relationships between different decision windows, which can lead to suboptimal performance. To address this issue, we propose a novel streaming decoding architecture, termed StreamAAD. In StreamAAD, decision windows are input to the network as a sequential stream and decoded in order, allowing for the modeling of inter-window relationships. Additionally, we employ a model ensemble strategy, achieving significant better performance than the baseline, ranking \textbf{First} in the challenge.

\end{abstract}

\section{Introduction}

Individuals with normal hearing can effectively track sounds of interest in complex auditory environments by modulating their attention \cite{cherry1953some}. However, hearing-impaired individuals encounter communication challenges in such scenarios even with the use of hearing aids \cite{bronkhorst2000cocktail}, as these devices cannot selectively enhance the sounds of interest. To address this issue, researchers have proposed the development of intelligent hearing aids that can detect users' focus of attention through neural signals and selectively amplify the desired speech \cite{hjortkjaer2024real, aroudi2020cognitive, han2024automatic}, thus achieving the goal of ``hearing what you want to hear." The technology of detecting auditory attention through neural signals is also known as auditory attention decoding (AAD) \cite{o2015attentional}.

AAD techniques can be broadly categorized into stimulus reconstruction (SR)-based methods and direct classification methods \cite{geirnaert2021electroencephalography, rotaru2024we}. The former utilizes specific algorithms to map Electroencephalography (EEG) signals to speech features, while the latter directly decodes the listener's attended spatial orientation from neural signals without relying on speech information, which is also known as spatial auditory attention decoding (Sp-AAD). Compared to SR-based methods, Sp-AAD does not depend on speech signals and can achieve decoding with shorter window lengths \cite{rotaru2024we,su2022stanet}, making it a focal point of recent research. The 14th International Symposium on Chinese Spoken Language Processing (ISCSLP 2024) hosts the Chinese Auditory Attention Decoding Challenge (Chinese AAD), encouraging researchers to explore Sp-AAD under various Chinese stimulus conditions. There are two tracks in this challenge: Single-Subject and Cross-Subject. We focus on Track 1, aiming to bulid an optimal decoder for each subject individually.

In recent years, researchers have made various efforts to achieve more accurate and faster auditory attention decoding. At the feature level, Qiu et al. \cite{qiu2024enhancing} utilized wavelet transform to extract the energy distribution across different frequency bands. Das et al. \cite{geirnaert2020fast} employed common spatial pattern features to capture spatial characteristics, while Jiang et al. \cite{jiang2022detecting} utilized multiband differential entropy to better capture the disorder degree of EEG signals. Regarding modeling techniques, neural networks have gained prominence due to their superior ability to handle nonlinear relationships. Vandecappelle et al. \cite{vandecappelle2021eeg} first used convolutional neural networks (CNNs) to capture spatial features in EEG signals. Cai et al. \cite{su2022stanet} applied temporal attention mechanisms to account for the brain's dynamic processing. Ni et al. \cite{Ni2024DBPNet} employed a dual-path network to model both time and frequency domains, and Xu et al. \cite{xu2024densenet} used 3D convolution to comprehensively explore the spatial distribution of EEG electrodes. Despite these advancements, a common issue persists: the decoding of different decision windows is isolated (referred to hereafter as the isolated decoding architecture). This isolation is detrimental to the decoding process.

Research indicates that EEG signal characteristics exhibit slow drifts over time \cite{rotaru2024we}, even within a single trial (specifically referring to auditory attention experiments lasting several minutes). In traditional isolated decoding architectures, EEG signals are divided into multiple independent decision windows, and the decoder is applied independently to each window during both training and testing phases. Such methods have two main drawbacks. First, they fails to model the temporal information across different windows, limiting its ability to account for EEG feature drift and restricting its performance to global invariant features. Second, due to the lack of correlation in decoding between adjacent windows, sudden decoding errors are more likely to occur, significantly affecting the user experience.

\begin{figure*}[t]
	\centering
	\includegraphics[width=150mm]{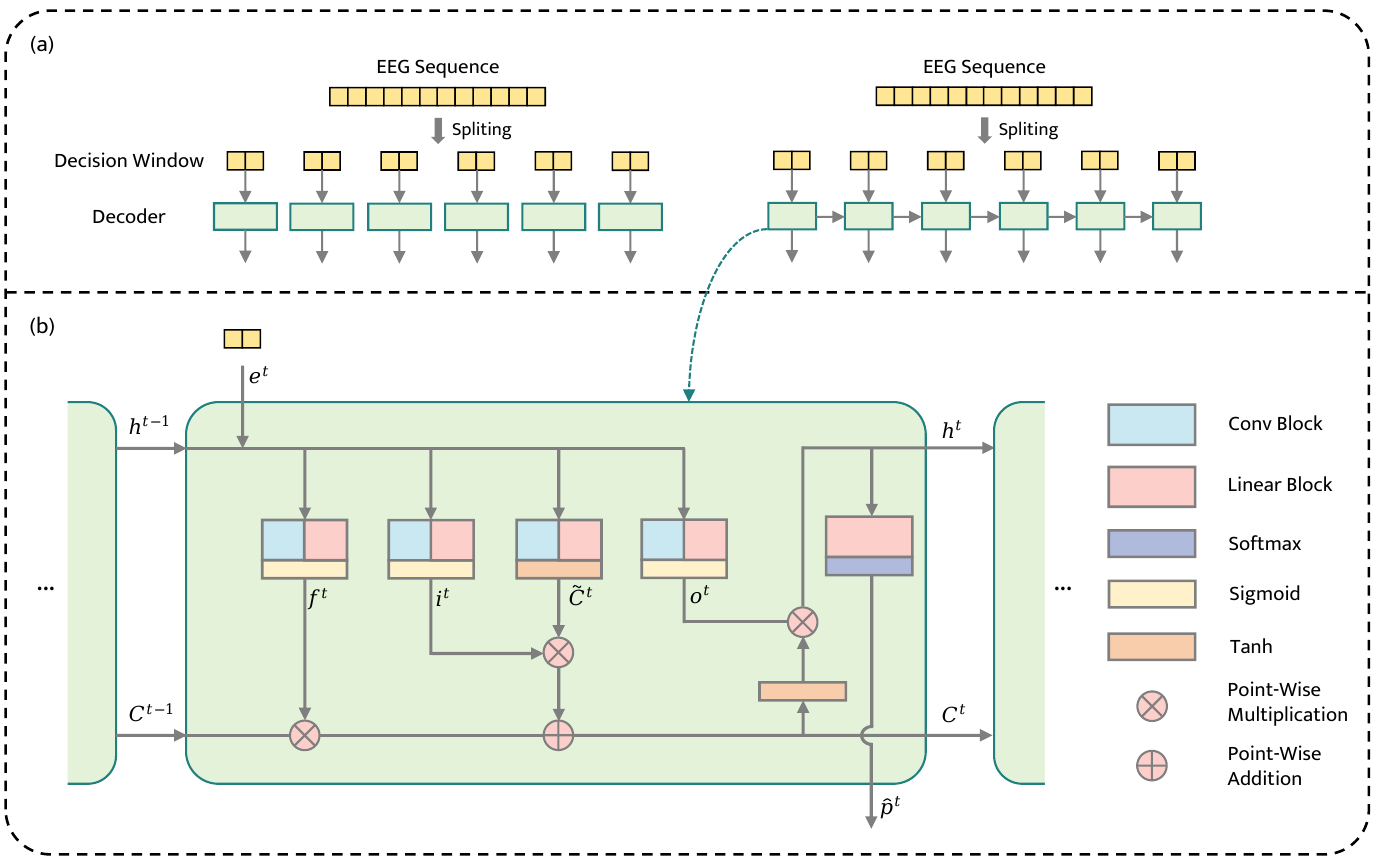}
	\caption{Overview of the proposed method. (a) Existing isolated window decoding architecture (left) and the proposed streaming decoding architecture, StreamAAD (right). (b) Detailed schematic of the decoder in StreamAAD.}
	\label{Figure1}
\end{figure*}

Therefore, we believe that modeling across different decision windows should not be completely isolated. Based on this, we propose a streaming decoding architecture termed StreamAAD, as illustrated in the right panel of Fig. \ref{Figure1}(a). During both training and testing phases, this architecture takes a sequence of decision windows as input and sequentially outputs the decoding result for each window. Compared to isolated window decoding architecture, the streaming decoding architecture also achieves real-time input and output. More importantly, the streaming architecture can model the relationships between decision windows, making it more robust to the inherent feature drift of EEG signals. Experiments demonstrate that the proposed StreamAAD outperforms existing algorithms. 

\section{Problem formulation}

The challenge involves a two-speaker scenario. Each subject, wearing EEG acquisition equipment, was presented with two simultaneous speech sources, one on each side (left and right). They were instructed to focus their attention on the speech from a specified direction. The EEG signals collected are preprocessed and segmented into decision windows. These windows are represented as $E \in \mathbb{R}^{T \times L \times C}$, with each decision window is denoted as $\bold{e}^t \in \mathbb{R}^{L \times C}$. Here, $T$ represents the number of decision windows, $L$ represents the length of each window, and $C$ is the number of channels, with $C=32$ in the provided dataset. The goal of the challenge is to develop a decoder $\mathfrak{D}(\cdot)$ that takes $E$ as input and decodes the direction of the subject's attention:
\begin{gather}
\hat{\bold{p}}=\frak{D}(E)
\end{gather}
Here, $\hat{\bold{p}} = [\hat{p}^1, \hat{p}^2, \ldots, \hat{p}^T]$ represents the decoding results for each decision window, where $\hat{p}^t \in \mathbb{R}^2$ indicates the probabilities of the target being on the left or right for the 
$t$-th window. For a given decision window, the decoding is considered correct if $\underset{j \in {0, 1}}{\operatorname{argmax}} \ \hat p[j] = y^t$, where $y^t$ is 0 or 1, representing the true label of the decision window.
\section{The proposed method}

Current Sp-AAD methods typically employ an isolated window decoding architecture, depicted on the left side of Fig. \ref{Figure1}(a). These methods divide an EEG sequence into multiple decision windows and decode each one independently. However, this approach fails to capture the slow time-varying characteristics of EEG signals. To address this limitation, we propose StreamAAD, as illustrated on the right side of Fig. \ref{Figure1}(a). Similar to recurrent neural network (RNN) \cite{medsker2001recurrent}, StreamAAD retains information after processing each decision window and uses it to decode the subsequent window, thus better accounting for the dynamic nature of EEG signals.

Long short-term memory network (LSTM) \cite{hochreiter1997long}, a type of RNN, is effective in modeling both long-term and short-term dependencies in sequences and is widely used in natural language processing and speech signal processing. Leveraging this, we employ an LSTM-like decoder in StreamAAD, as shown in Fig. \ref{Figure1}(b). Different from a normal LSTM, in StreamAAD, information is transmitted at the decision window level rather than the sampling point level. Specifically, given an input sequence of decision windows $E=[\bold{e}^1,\bold{e}^2,...,\bold{e}^T]$, at the $t$-th time step, the forget gate, input gate, output gate, and cell gate are calculated using the current input decision window $\bold{e}^t$ and the long-term and short-term memory contents $C^{t-1}$ and $h^{t-1}$ from the previous step:
\begin{gather}
	f^t,i^t,o^t=\text{Sigmoid}(\text{Conv}(\bold{e}^t)+\text{Linear}(h^{t-1}))\\
	\tilde{C}^t=\text{Tanh}(\text{Conv}(\bold{e}^t)+\text{Linear}(h^{t-1}))
\end{gather}
Here, $f^t$, $i^t$, $o^t$, and $\tilde{C}^t$ are hidden vectors of length $D$. Since the shapes of the memory contents and the input decision window differ, we employ different computation methods for each. For the input $\bold{e}^t$, we use a convolutional block (Conv Block) to extract features, which includes a 1D convolution layer, a ReLU \cite{nair2010rectified} activation function, a global average pooling layer along the temporal dimension, and a LayerNorm \cite{ba2016layer} layer. For the short-term memory content $h^{t-1}$, we use a linear block consisting of a fully connected layer and a ReLU activation function. Once these gates are obtained, the information passed to the next time step and the decoding output are computed as follows:
\begin{gather}
C^t=f^t\cdot C^{t-1}+i^t\cdot \tilde{C}^{t}\\
h^t=\text{Tanh}(C^t)\cdot o^t\\
\hat{p}^t=\text{Softmax}(\text{Linear}(h^t))
\end{gather}
In the context, $C^t$ and $h^t$ represent the long-term and short-term states at the current step, respectively. These states capture the long-term time-varying characteristics and the short-term attention features of the EEG signals. The model is trained in an end-to-end way using a cross-entropy loss function:
\begin{gather}
	L= -[y \log \hat p + (1-y)\log (1-\hat p)]   
\end{gather}

It is important to note that we use a simple convolutional neural network as the basic computational unit in StreamAAD. However, the proposed StreamAAD is a versatile streaming decoding architecture, allowing its internal computational unit to be easily replaced with more complex neural networks.

\section{Experimental setups}
\subsection{Data specifications}
\label{Section4.1}
We utilize the official dataset (raw) \cite{Zhang2024AudioVideoEEG} for model training and validation. This dataset includes EEG signals from eight subjects (SubA-H) collected during auditory attention experiments in both audio-only and audio-video scenarios. Each subject underwent 16 trials per scenario, with each trial lasting about 150 seconds.

The original data, sampled at 1k Hz, is first downsampled to 128 Hz. We then apply high-pass and low-pass filters with cutoff frequencies of 1 Hz and 45 Hz, respectively. The values of the processed EEG signal are very small, indicating they are measured in volts. Therefore, we multiply all data by $10^6$ to convert the values to microvolts. As our decoding approach is end-to-end, we do not perform artifact removal.
\subsection{Training setups}
For each subject, we further divide each trial in the official training set (about 136 second each trial) into a training set and a validation set at an 8:1 ratio, consistent with the organizers' method. This ensures the validation set size matches that of the official test set. For the proposed method, we segment the data into 5-second segments, which are then further divided into sequences of nine consecutive decision windows ($E\in\mathbb{R}^{9\times128\times32}$) using a sliding window of 1.0 second with a  0.5-second stride. For contrasive methods, we apply the sliding window directly to each trial.

In the Conv Block of the proposed StreamAAD, we use 1-D convolution with a kernel size of 7 and a filter number of 32. The length of hidden vector  is set to match the number of filters, thus $D=32$. Each model is trained separately for each subject, utilizing both audio-only and audio-video data. We employ the Adam optimizer \cite{kingma2014adam} for training, with a learning rate of 0.001 and a batch size of 8, and we train each model for 150 epochs. The $Wilcoxon$ signed-rank test \cite{wilcoxon1992individual} is employed to compare the performance of different methods.

\section{Results and analysis}
\subsection{Main results}
\begin{figure*}[t]
	
	\centering
	\includegraphics[width=167mm]{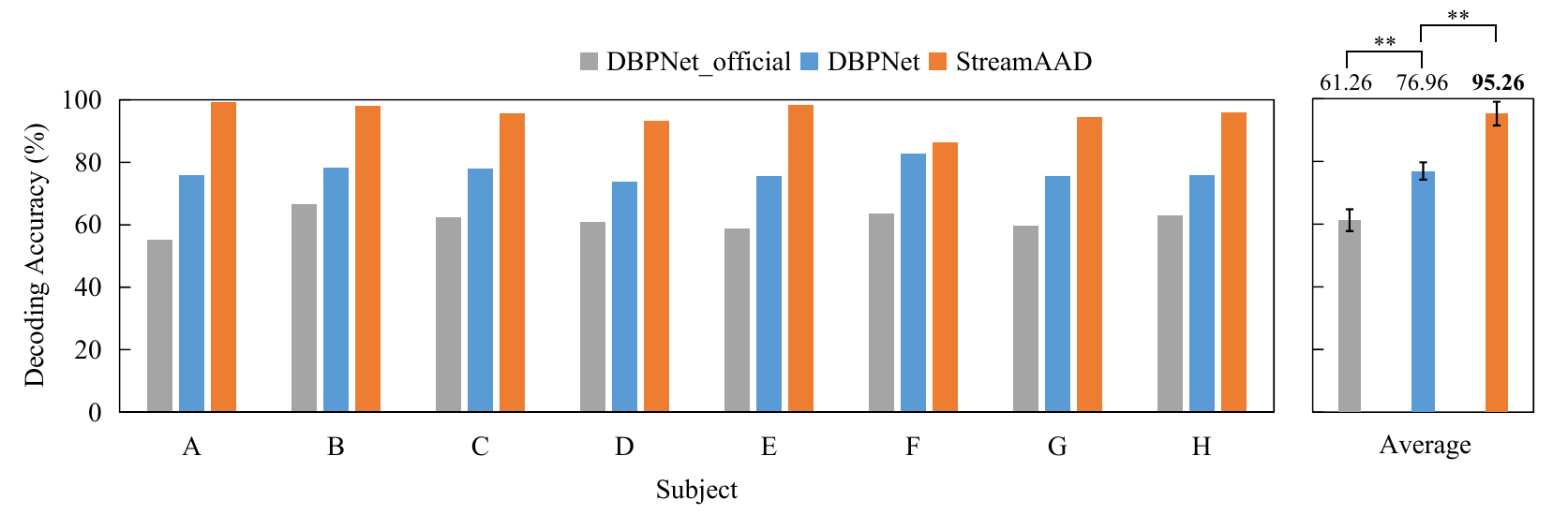}
	\caption{Decoding accuracy of the proposed method compared to other methods across different subjects (left) and the average decoding accuracy across all subjects (right). The results of DBPNet\_official are from the organizer (**: $p\textless 0.01$).}
	\label{Figure2}
\end{figure*}

We first report the experimental results on the test set. For comparison, we also report the performance of the challenge baseline, DBPNet \cite{Ni2024DBPNet}, a recently proposed model with superior performance. As shown in Fig. \ref{Figure2}, the average decoding accuracy of DBPNet across the eight subjects is 76.96\% (SD: 2.75\%), while StreamAAD achieves an accuracy of 95.26\% (SD: 3.80\%), outperforming DBPNet by 18.30\% ($p\textless 0.01$). This demonstrates the efficiency of the proposed algorithm. Although DBPNet jointly models temporal and spectral features, it still employs an isolated window decoding architecture, failing to model inter-window relationships in EEG signals. In contrast, despite using a simple Conv Block as the backbone, the streaming architecture of StreamAAD captures the time-varying characteristics of EEG, leading to better attention detection. We also compare the parameter count and computational cost of the two models, as shown in Table \ref{Table1}. The proposed method requires less than one-twentieth of the parameters and the number of multiplication and accumulation operations (MACs) compared to DBPNet, and it also uses less memory. This further underscores the importance of modeling the relationships between decision windows.

\begin{figure*}[t]
	\centering
	\includegraphics[width=167mm]{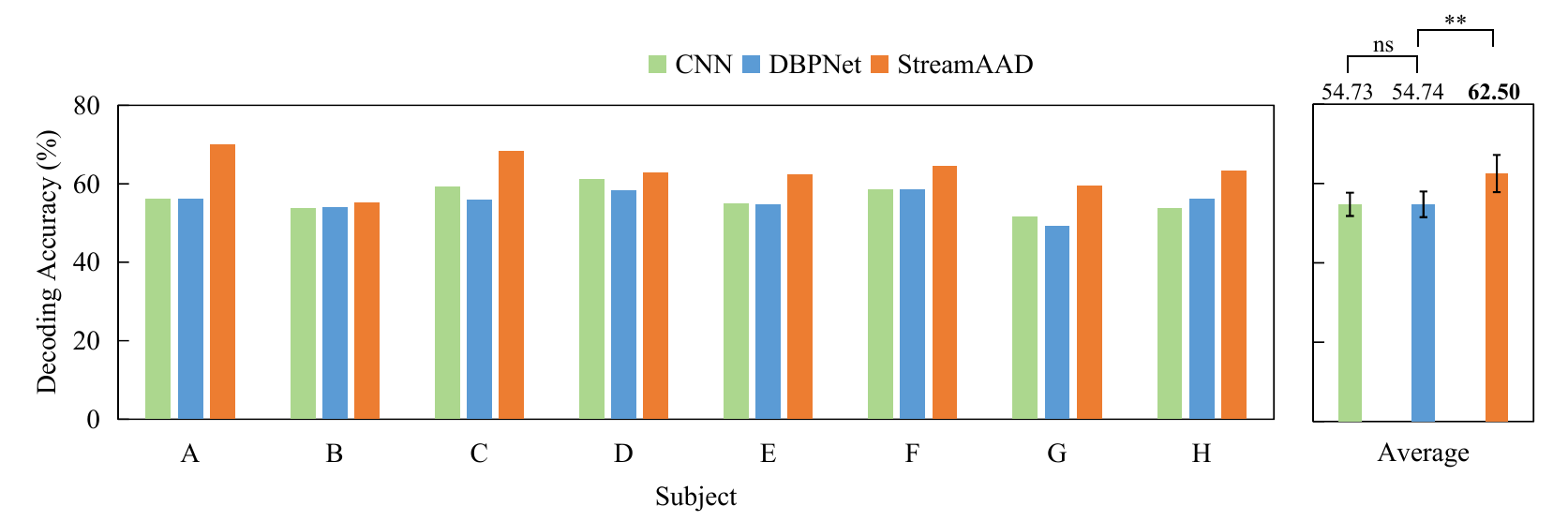}
	\caption{Decoding accuracy of the proposed method compared to other methods across different subjects (left) and the average decoding accuracy across all subjects (right) using cross-trial data partitioning strategy (**: $p\textless 0.01$, ns: not significant).}
	\label{Figure3}
\end{figure*}

\begin{table}[t]
	\caption{Comparison of StreamAAD and DBPNet in terms of parameter count, computational cost, and memory usage, using 5-second data segments as an example.}
	\label{Table1}
	\begin{tabular}{@{}cccc@{}}
		\toprule
		Model     & \#Params (K) & MACs (M) & Memory (MB) \\ \midrule
		DBPNet    & 870.8        & 823.9    & 484.0       \\
		StreamAAD & 33.3         & 31.5     & 348.0       \\ \bottomrule
	\end{tabular}
\end{table}

Additionally, we present the test results of DBPNet provided by the challenge organizer, shown by the gray bars in Fig. \ref{Figure2} (DBPNet\_official). The average accuracy is 61.26\% (SD: 3.47\%), which is lower than our tested accuracy. We believe this discrepancy may be due to the official model being trained on data measured in volts, which has very small values, making the model harder to converge.
\subsection{Results under cross-trial data partitioning strategy}
In this track, all test data has corresponding data from the same trials in the training set, following a within-trial data partitioning strategy. However, recent research \cite{rotaru2024we,qiu2024enhancing, xu2024beware} indicates that EEG from the same trial may exhibit similar features, known as trial-specific features. This can cause the network to learn these features and infer trial information rather than actually decoding attention, potentially leading to inflated results. In contrast, a cross-trial data partitioning strategy, where test data comes from different trials with the training data, prevents the model from relying on trial-specific features and offers a \textbf{MORE OBJECTIVE} evaluation of model performance.

Therefore, we also evaluate the cross-trial performance of the proposed model. To prevent the model from learning trial-specific features, we retain only the alpha and beta frequency bands, which are closely related to spatial auditory attention \cite{vandecappelle2021eeg, xu2024beware, deng2020topographic}, by applying a 7-30 Hz band-pass filter to the data. Since the trial order of the official test data was shuffled and the trial indices are unknown, we use the first 10 trials from the official training set for each subject as training data, the last four trials as testing data, and the remainder as validation data.

For comparison, we also include an additional CNN model. The CNN model is composed of a Conv Block and a Linear Block, both identical to those in StreamAAD. As shown in Fig. \ref{Figure3}, both StreamAAD and DBPNet experience a significant drop in decoding accuracy when employing the cross-trial data partitioning strategy compared to the within-trial strategy, consistent with previous studies \cite{rotaru2024we, qiu2024enhancing, xu2024beware}. However, StreamAAD still demonstrates a notable advantage over DBPNet ($p\textless 0.01$). Moreover, StreamAAD also outperforms CNN ($p\textless 0.01$), further illustrating the superiority of the proposed streaming decoding architecture.
\subsection{Submitted results}
\begin{table}[]
	\centering
	\caption{Decoding accuracy using different data and ensemble strategies, with the final submitted result highlighted in bold.}
	\label{Table2}
	\tabcolsep=0.09cm
\begin{tabular}{@{}c|c|c@{}}
	\toprule
	Data for Training & Ensemble Strategy            & Decoding Accuracy (\%) \\ \midrule
	Without Val Set   & None                           & 95.26$\pm$3.85                  \\
	Val Set Included  & Across Epochs       & 97.74$\pm$1.56                   \\
	Val Set Included  & Across Runs & \textbf{99.06$\pm$0.75}                  \\ \bottomrule
\end{tabular}
\end{table}

To further enhance the model's performance, we employ two strategies in the final submission. First, we use all the provided data, including the validation set mentioned in Section \ref{Section4.1}, as training data. Without a validation set, we adopt an ensemble strategy \cite{sagi2018ensemble} by averaging the output probabilities from the last ten epochs of training: $\hat{p}^t_{ensem\_1} = \displaystyle{\frac{1}{10}\sum_{i=0}^{9}\hat{p}^t_{N_{epoch}-i}}$,
where $N_{epoch}=150$ is the total number of training epochs. Second, to further boost performance, we train the model 100 times with different random seeds. The final result is the average of these 100 outputs:
$\hat{p}^t_{final} =  \displaystyle{\frac{1}{100}\sum_{i=1}^{100}\hat{p}^t_{ensem\_i}}$.
The results using different data and strategies are shown in Table \ref{Table2}.
\section{Conclusion}
In this paper, we present our approach for the Chinese AAD Challenge at ISCSLP 2024. We propose a streaming auditory attention decoding architecture to model the relationships of between decision windows. Experimental results demonstrate that the proposed StreamAAD outperforms existing isolated window decoding architectures, achieving superior decoding performance with fewer computational resources under various data partitioning strategies. By employing an ensemble strategy, we achieve a higher decoding accuracy on the official test set, ranking first in the Track 1 of Chinese AAD.
\section{Acknowledgements}
This research is partially supported by STI 2030—Major Projects (No. 2021ZD0201503), the National Natural Science Foundation of China (No. 12104483).

\bibliographystyle{IEEEtran}

\bibliography{mybib}


\end{document}